\documentclass[11pt,a4paper]{article}
\usepackage{jheppub}
\usepackage{xcolor}
\usepackage{amsmath}
\usepackage{physics}
\usepackage{graphicx}
\usepackage{svg}
\graphicspath{{./images}}
\usepackage{subcaption}
\usepackage{array}
\usepackage{tikz}
\usepackage{url}
\usepackage{multirow}
\usepackage{tabularx}
\newcolumntype{Y}{>{\centering\arraybackslash}X}
\usepackage{makecell}
\newcommand{\eq}[1]{eq.~(\ref{#1})}

\usepackage{hyperref}

\usepackage{hyperref}

\begin{document}

\title{\boldmath Cosmological selection of a small weak scale from     large vacuum energy: a minimal approach}


\definecolor{lime}{HTML}{A6CE39}
\DeclareRobustCommand{\orcidicon}{\hspace{-1mm}
	\begin{tikzpicture}
	\draw[lime, fill=lime] (0,0) 
	circle [radius=0.12] 
	node[white] {{\fontfamily{qag}\selectfont \tiny \,ID}};
	\draw[white, fill=white] (-0.0525,0.095) 
	circle [radius=0.007];
	\end{tikzpicture}
	\hspace{-3mm}
}

\foreach \x in {A, ..., Z}{\expandafter\xdef\csname orcid\x\endcsname{\noexpand\href{https://orcid.org/\csname orcidauthor\x\endcsname}
		{\noexpand\orcidicon}}
}

\newcommand{\orcidauthorA}{0009-0000-2346-2273}
\newcommand{\orcidauthorB}{0000-0003-2323-3950}
\newcommand{\orcidauthorC}{0000-0003-4398-4698}


\preprint{TIFR/TH/24-14}

\author[a]{Susobhan Chattopadhyay\orcidA{}}
\author[a]{Dibya~S.~Chattopadhyay\orcidB{}}
\author[a]{Rick S. Gupta\orcidC{}}

\affiliation[a]{Tata Institute of Fundamental Research, Homi Bhabha Road, Colaba, Mumbai 400005, India}

\emailAdd{susobhan.chattopadhyay@tifr.res.in}
\emailAdd{d.s.chattopadhyay@theory.tifr.res.in}
\emailAdd{rsgupta@theory.tifr.res.in}

\abstract{ 
We present a minimal cosmological solution to the hierarchy problem. Our model consists of a light pseudoscalar and an extra Higgs doublet in addition to the field content of the Standard Model. We consider a landscape of vacua with varying values of the electroweak vacuum expectation value (VEV). The vacuum energy in our model peaks in a region of the landscape where the electroweak VEV is non-zero and much smaller than the cutoff. During inflation, due to exponential expansion, such regions of the landscape with maximal vacuum energy, dominate the universe in volume, thus explaining the smallness of the electroweak scale with respect to the cutoff. The pseudoscalar potential in our model is that of a completely generic pseudogoldstone boson---not requiring the clockwork mechanism---and its field value never exceeds its decay constant or the Planck scale.  Our mechanism is robust to the variation of other model parameters in the landscape along with the electroweak VEV. It also predicts a precise and falsifiable relationship between the masses and couplings of the different Higgs boson mass-eigenstates. Moreover, the pseudoscalar in our model can account for the observed dark matter relic density.}

\maketitle

\flushbottom
\section{Introduction}
\label{sec:intro}
The hierarchy problem has been one of the central themes of particle physics research in the last few decades. Most of the approaches to solve the hierarchy problem involve embedding the Standard Model into a larger framework where some new symmetry---such as supersymmetry or a shift symmetry---is recovered in the limit, $\mu^2\to 0$, $\mu^2$ being the Higgs mass squared parameter. The Higgs mass in these theories is thus related to the scale of breaking of this new symmetry which results in the prediction of new particles at the TeV scale such as superpartners or composite states. There is thus a growing tension between the predictions of these models and the lack of any evidence for new physics at the LHC.

This situation has led to the emergence of a new approach to the hierarchy problem where the, $\mu^2=0$, point is special---not from the point of view of symmetry---but because it separates two different phases, namely, the electroweak breaking and preserving phases. Models using this approach~\cite{Dvali1, Dvali2, gkr, nnat,  Geller, Cheung, GiudiceK,Strumia, Csaki,trigger, Giudicesol,  Tito1,Tito2,Csaki2, Mats} propose mechanisms for the Higgs sector to spontaneously evolve to a near-critical state, making them somewhat reminiscent of systems exhibiting self-organized criticality~\cite{soc, Giudice}. They use the electroweak vacuum expectation value (VEV) to trigger some cosmological dynamics that provides a non-anthropic mechanism to select a small but negative Higgs mass squared from a landscape of possible values, $-\Lambda^2\lesssim \mu^2 \lesssim \Lambda^2$,    $\Lambda$ being the cutoff. This landscape can either be physically realized by a variation of $\mu^2$ over causally disconnected patches during inflation or, as in relaxion models~\cite{gkr},   can be scanned only in time, for instance, by a slowly rolling scalar field. Unlike the anthropic approach~\cite{Weinberg, Agrawal}, these models provide clear testable signatures.

The central idea of our work is the observation that there is already a physical quantity in the Standard Model---namely the vacuum energy contribution from the Higgs potential---that behaves in a special way at the boundary between the electroweak symmetry broken and unbroken phases (see also Ref.~\cite{Cheung, Giudicesol}). In the unbroken phase,  the vacuum energy contribution is constant and independent of the Higgs potential parameters. In the broken phase, on the other hand,  it drops monotonically with the electroweak VEV as the Higgs minimum becomes deeper. We propose a model where the Standard Model (SM) Higgs sector is promoted to a two Higgs doublet model (2HDM) with an additional light pseudoscalar, $\phi$. The vacuum energy of the 2HDM-$\phi$ system decreases for large electroweak VEVs---a feature our model inherits from the SM. Unlike the SM, however,  the vacuum energy contribution peaks at a non-zero value of the electroweak VEV that is much smaller than the cutoff.

\begin{figure}[t]
		\centering
		\includegraphics[width=0.9\linewidth]{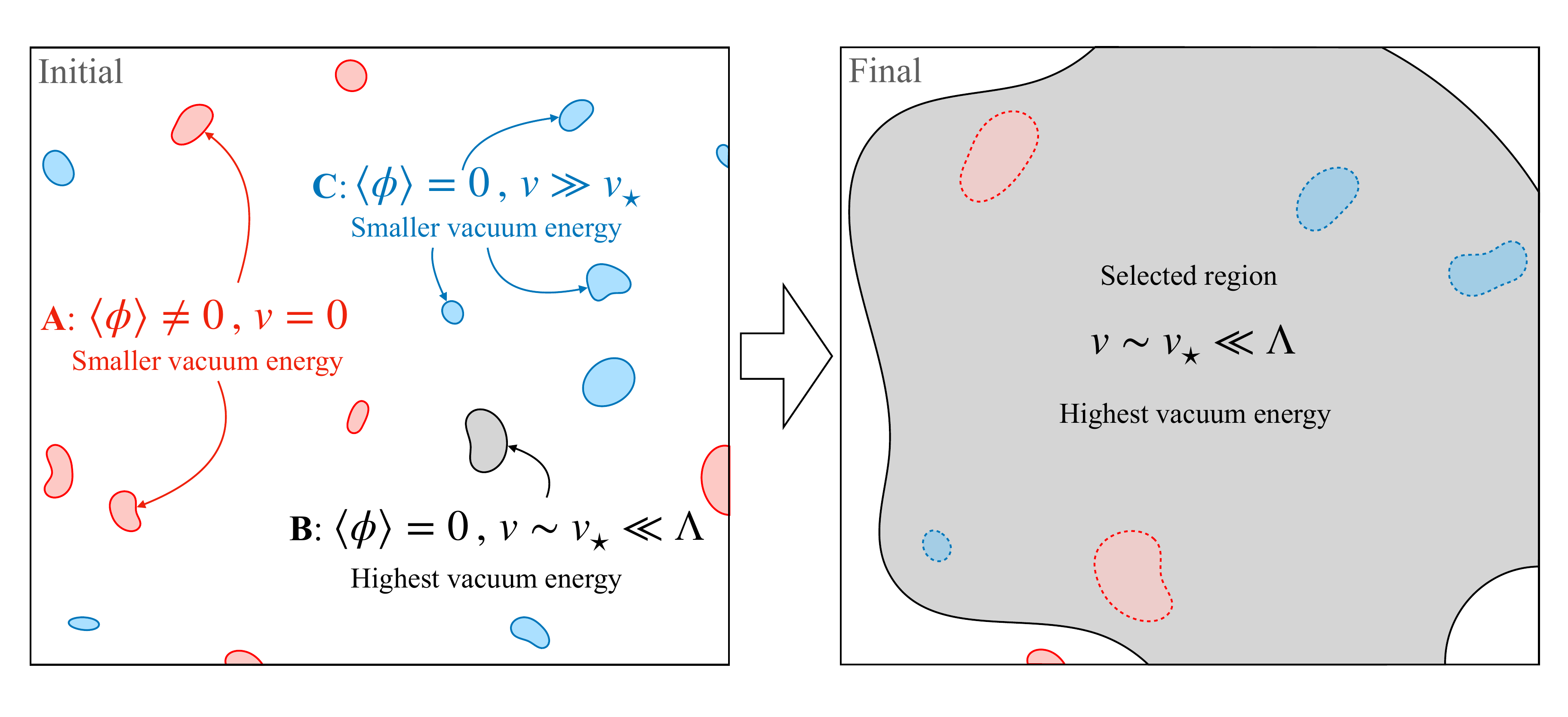}
		\caption{A schematic description of our mechanism. In the left panel, we show casually disconnected patches with different values of the electroweak VEV, $v$, in an eternally inflating universe. The vacuum energy is maximum in the gray patches labeled B where   $v=v_\star \ll \Lambda$. These gray regions then expand exponentially faster than other patches.  As shown in the right panel, this eventually results in these patches---with a hierarchically small electroweak scale---dominating the volume of the universe.  For the red patches where the electroweak VEV vanishes (labeled A) and the blue patches where it is much larger than the observed value $v_\star$ (labeled C) the vacuum energy is smaller than that in the gray patches by construction. The red and blue patches with dotted outlines, within the grey region in the right panel, represent such lower energy vacua generated by tunneling from the maximal energy vacuum (see Sec.~\ref{measure}). We show the potential for the pseudoscalar and the Higgs doublets in our model corresponding to the three possibilities (A, B, and C) in Fig.~\ref{fig:2}.}
		\label{fig:1}
	\end{figure}

We assume a landscape of values for the parameters of the 2HDM-$\phi$-potential that is realized in an eternally inflating `multiverse'.  We then utilize the observation made by Ref.~\cite{Geller}  (and further developed in  Ref.~\cite{Cheung,Giudicesol, Strumia}) that regions of the multiverse with maximal vacuum energy---that in our model also have a large hierarchy between the electroweak scale and the cutoff---will expand at an exponentially faster rate compared to other regions and will thus eventually dominate in volume. A schematic description of our mechanism has been presented in Fig.~\ref{fig:1}. A possible concern in all models that utilize the idea of Ref.~\cite{Geller},  is the implicit use of a volume based measure for the multiverse; this can be problematic because of the so-called youngness paradox~\cite{guthYP}. As we will show, however,  with an appropriate choice of the measure, our mechanism can explain the Higgs mass hierarchy while not running into any such paradoxes. 

Our model has several attractive features: (1) the $\phi$-potential   in our model is that of a generic light scalar, for eg. a pseudo-Nambu goldstone boson (PNGB), and does not require a clockwork-like mechanism; (2)  the $\phi$-field value never exceeds its `decay constant', $f$, let alone the Planck scale;  (3) unlike the anthropic argument for a weak scale~\cite{Agrawal}, our mechanism does not restrict the variation of other model parameters as the electroweak VEV is varied; (4)  the number of e-folds in the slow-roll phase of inflation does not need to be very large; (5)  there is a precise, falsifiable prediction that analytically relates different masses and couplings of the 2HDM sector; and finally, {(6)} the pseudoscalar, $\phi$, can account for the observed dark matter density via the misalignment mechanism.

The plan of the paper is as follows. In Sec.~\ref{sec: model} we describe our mechanism in detail. In Sec.~\ref{sec:2hdm} we discuss the phenomenology of the 2HDM sector whereas in Sec.~\ref{phipheno} we discuss the phenomenology of the $\phi$-field and the possibility that it can account for the observed dark matter relic density. Finally, we make some concluding remarks in Sec.~\ref{conclusions}.

\section{A small weak scale from large vacuum energy}\label{sec: model}
\subsection{Model}

\label{higgssec}

We consider a modified Higgs sector consisting of two  Higgs doublets, $H_1$ and $H_2$, and a pseudoscalar,  $\phi$.  We assume that the fundamental theory has a landscape of vacua with varying values of the scalar potential parameters and thus varying values for the electroweak VEV. We then consider an eternally inflating universe where these different vacuum configurations are realized in different causally disconnected patches of the universe.

The pseudoscalar in our model  has the following  potential,
\begin{equation}
		V_\phi(\phi) = \mu_\phi^2 f^2 \left(-\frac{1}{2}\left(\frac{\phi}{f}\right)^2 + \lambda_\phi\left(\frac{\phi}{f}\right)^4+\dots \right)
		\label{vphi}
	\end{equation}
where $ \mu^2_\phi \ll f^2$ and $\lambda_\phi\sim \mathcal{O}(1)$ are both positive. The field value of $\phi$ in our model never exceeds $f$ so that the above equation presents a consistent effective field theory (EFT) expansion. A potential  of the above form  can arise if  $\phi$ is the PNGB of a global symmetry that is spontaneously broken at a scale $f$,  with,  $M\sim \sqrt{\mu_\phi f}$, being the  scale at which the symmetry is broken explicitly or due to an anomaly.\footnote{For a PNGB the above potential would arise as a linear combination of periodic terms such as $M^4 \cos \frac{\phi}{f}, M^4 \cos \frac{2\phi}{f}$, etc.} 

The `mexican-hat' potential, $V_\phi$,  is minimized at a non-zero value of $\phi$ and gives a negative contribution to the vacuum energy. The central idea of our mechanism is that the electroweak symmetry breaking can trigger a change in $V_\phi$, from a mexican-hat potential to one which is minimised at the origin.
This raises the vacuum energy of patches where the electroweak VEV is above some threshold value required to cause such a change in $V_\phi$.  As in the SM, very large electroweak VEVs,  give a large negative vacuum energy contribution to the potential.  The net effect of these two factors is that the vacuum energy in our model peaks in patches with a finite but small electroweak VEV. Away from this value, we get a large negative contribution either due to the electroweak or   $\phi$ VEV.

The simplest possibility to realize this would be the addition of a `trigger term'~\cite{trigger}  to our potential of the form $V_T\sim |H|^2 \phi^2$. This term can give the desired positive contribution to the pseudoscalar mass. However, such a term is problematic as we can close the Higgs loop to obtain a  quadratically  divergent term,
\begin{eqnarray}
 |H|^2 \phi^2 \to  \frac{\Lambda^2}{16 \pi^2} \phi^2.
 \label{closeloop}
\end{eqnarray}
This would mean that the change in the $V(\phi)$ that we wish to trigger by the electroweak VEV can already be caused by closing the Higgs loop. The loop-induced effect is in fact bigger than that induced by the Higgs VEV unless, $\Lambda \lesssim 4 \pi v$,  which kills any hope of solving the hierarchy problem up to high scales.

  While cosmological selection models employ diverse mechanisms to select the weak scale, many of them require a term of the form $\phi^n H^\dagger H$ in the potential that is triggered by the electroweak VEV. All such models, therefore, face the issue related to the closing of the Higgs loop, described in the previous paragraph. For our model,  we will utilize the minimal solution to this problem provided by Ref.~\cite{trigger}, where it was proposed that the operator $H_1^\dagger H_2$ in a two Higgs doublet model (2HDM)  can act as the electroweak trigger.\footnote{An alternative resolution to this issue can arise if the loop in \eq{closeloop} is cutoff at a scale much smaller than the Higgs cutoff, $\Lambda$ (see for eg. Ref.~\cite{gkr, pomarol, Gupta1, Gupta2,Gupta3}).} For our model, we thus consider the following trigger term,  
\begin{equation}
V_T(\phi, H_1, H_2)=\frac{\mu_\phi^2 f^2}{2}\left(\kappa \left(\frac{H_1^\dagger H_2}{\mu_\phi f}\right) + h.c.\right)\left(\frac{\phi}{f}\right)^2,
\label{trigger}
 \end{equation}
where  perturbativity requires,
\begin{equation}
\kappa^2 \lesssim  16 \pi^2  \lambda_\phi. \; 
\label{upperB}
\end{equation}
With this trigger term, it is clearly not possible to generate a one-loop contribution like the one in \eq{closeloop}. We also  impose a $\mathbb{Z}_2$ symmetry, $H_1 \to -H_1$ which leads to the following 2HDM potential, 
\begin{align}
  \label{2hdmpot}
	V_{\text{2HDM}}(H_1, H_2) = & \mu_1^2 H_1^\dagger H_1 + \mu_2^2 H_2^\dagger H_2 + \lambda_1 (H_1^\dagger H_1)^2 + \lambda_2 (H_2^\dagger H_2)^2 + \lambda_3 (H_1^\dagger H_1)(H_2^\dagger H_2) \nonumber \\
	& + \lambda_4 (H_2^\dagger H_1)(H_1^\dagger H_2) + \frac{1}{2}\left(\lambda_5 (H_1^\dagger H_2)^2 + \lambda_5^* (H_2^\dagger H_1)^2\right).
\end{align}

Note that apart from $\kappa$ and $\lambda_5$, all of the coupling above is real by hermiticity. Furthermore, we will assume a real and positive  $\kappa$,  as any accompanying phase can always be removed by rotating $H_2$.  As far as $\lambda_5$ is concerned, we will rewrite it as,   $\lambda_5= \hat{\lambda}_5 e^{i\alpha}$, where $\hat{\lambda}_5<0$.

As discussed in Ref.~\ref{trigger}, the  $\mathbb{Z}_2$ symmetry imposed on $V_{\text{2HDM}}$ is essential to prevent  2HDM potential terms like $(H_1^\dagger H_1)(H_1^\dagger H_2)$, that can, at the two-loop level, generate a contribution like the one in \eq{closeloop} by closing the Higgs doublets in \eq{trigger}. The $\mathbb{Z}_2$ symmetry is actually approximate as the trigger term in \eq{trigger} breaks it. In fact, this breaking reintroduces    contribution to the $\phi$-mass term at the two loop level, \footnote{Note that  the term, $\left( \kappa  \frac{\mu_\phi}{2 f} \frac{\mu_\phi^2}{16 \pi^2} H_1^\dagger H_2 +h.c.\right)$ is similarly  generated by closing the $\phi$-loop in the trigger term. The coefficient of the term is again completely negligible and can be safely ignored for the rest of the analysis.}
\begin{equation}
\Delta V_\phi^{2-loop}\sim\kappa^2 \frac{\mu_\phi^2}{f^2} \frac{\mu_\phi^2}{(16 \pi^2)^2} \phi^2.
\label{2loop}
\end{equation}
Given the upper bound on $\kappa$ in \eq{upperB}, one can check that this contribution is much smaller than that obtained by substituting the electroweak VEV in \eq{trigger}---thus not spoiling the effectiveness of the trigger term.

 Collecting all the different terms we thus get the full potential of the Higgs sector,
\begin{equation}
    V_H (\phi, H_1, H_2)= V_\phi (\phi) +V_T (\phi, H_1, H_2)+V_{2HDM}(H_1, H_2).
    \label{vh}
\end{equation}



\subsection{Cosmological set-up and the landscape}
\label{land}

We  assume that the Higgs sector described in Sec.~\ref{higgssec}  is completely decoupled from the inflaton sector and that the total vacuum energy in a particular minima in the landscape is given by,  
\begin{equation}
    {\cal VE}^{(ijk)}={\cal VE_H}(\{\alpha^{(i)}_{H}\}, P(\phi, H_1, H_2))+{\cal VE}_\chi(\{\beta^{(j}_{\chi}\}, P(\chi)))+(\Lambda^{(k)}_{cc})^4,
    \label{vace}
\end{equation}
where $\chi$ is the inflaton field. The first term denotes the contribution from $V_H$ in \eq{vh}, which vanishes if all the Higgs sector fields are localized at the origin,  $\phi= H_1=H_2=0$; the second term denotes the contribution from the inflaton potential which vanishes at $\chi=0$; and the third term is the contribution of the cosmological constant.  Here $\{\alpha^{(i)}_{H}\}$ denotes a particular choice of  Higgs sector parameters, namely $\{\mu_1^2, \mu_2^2, \lambda_{1}, \lambda_2, \lambda_3, \lambda_4, \lambda_5, \kappa, |\mu_\phi^2|, \lambda_\phi\}$; $\{\beta^{(j}_{\chi}\}$ denotes a choice of parameters for the unspecified inflaton potential; and,   $\Lambda^{(j)}_{cc}$, is a particular choice for the cosmological constant. The relevant parameters in our theory vary in the range,  $-\Lambda^2 <\mu_i^2 <\Lambda^2$ and $-\Lambda^4 <\Lambda_{cc}^4 <\Lambda^4$,  across different vacua of the landscape. The scale up to which the hierarchy problem is solved in our model is thus,  $\Lambda$.   Note that the mexican hat nature of $V_\phi$ is crucial to our mechanism and the sign of $\mu_\phi^2$ in \eq{vphi}   is thus assumed to be positive across the landscape. We make no assumptions about the origin of the landscape which can arise either from string theory~\cite{kklt1,kklt2,susskind}  or from a sector at lower energies having multiple scalars (see for eg. Ref~\cite{trigger,Tito1}).\footnote{Note that, for our selection mechanism to work, we require an exponentially large number of vacua to ensure that the electroweak VEV is scanned finely in the landscape. This is not, however, a severe issue because the number of vacua in the landscape also grows exponentially with the number of scalars.}

Our assumption that the Higgs sector and inflaton sector are decoupled from each other is an important one that we now discuss in more detail. In particular we assume (1) that a transition from one vacuum to another in the landscape that alters the Higgs sector parameters,  $\{\alpha^{(i)}_{H}\}$, does not change the inflaton sector parameters, $\{\beta^{(j)}_{H}\}$ and vice versa; and (2) any coupling between the inflaton, $\chi$ and the Higgs sector fields is highly suppressed.

In \eq{vace}, $P(\chi)$ and $P(\phi, H_1, H_2)$,  denote the volume-weighted distribution functions for the scalar field values during inflation. We will discuss more about the inflationary dynamics and $P(\chi)$ in Sec.~\ref{measure}.  For the Higgs sector fields, $P(\phi, H_1, H_2)$ can be computed by solving the volume-weighted Fokker Planck equation (FPV). As we discuss in detail in App.~\ref{FPV}, a careful analysis of the FPV equation shows that the probability distribution function is sharply peaked at the classical minima of $V_H$ in \eq{vh}  if the following conditions hold, 
\begin{eqnarray}\label{cbeatsq}
\frac{H_I^4}{v_\star^4} \ll 1, \qquad \frac{H_I^4}{\mu^2_{\phi} f^2} \ll1, \qquad \frac{f^2}{M_{pl}^2} \ll 1,
 \end{eqnarray}
where $H_I$ is the Hubble scale during inflation and $v_\star=246$ GeV is the observed electroweak VEV.  We will assume this to be the case in our model. This assumption simplifies the computation of the vacuum energy contribution from the Higgs sector, ${\cal VE_H}$.  In particular, if, for  a particular choice of parameters $\{\alpha^{(i)}_{H}\}$,  $V_H$ has a single minimum, we obtain, 
\begin{equation}\label{single}
{\cal VE_H}(\{\alpha^{(i)}_{H}\})= V_H(\{\alpha^{(i)}_{H}\} \, ,\, \phi_{0}\, ,\, H_{1,0} \, , \, H_{2,0})
\end{equation}
where $(\phi_{0} \, , \, H_{1,0} \, ,\,  H_{2,0})$ is the position of the minimum.

We will show in Sec.~\ref{measure} that inflationary dynamics would result in the multiverse being dominated by the vacuum state where each of the terms in \eq{vace}---and in particular the Higgs contribution, ${\cal V_H}$---is maximized. We expect this maximal vacuum energy to be of the order,
\begin{equation}\label{cutoff}
H_I^2 M_{pl}^2 \sim \Lambda^4,
\end{equation}
as the maximal  value of,  $\Lambda_{cc}$, is of the order of the cutoff, $\Lambda$. The first equation of \eq{cbeatsq} thus implies an upper bound on the cutoff,
\begin{equation}\label{cutoffb}
    \Lambda\sim 10^{10} {\rm GeV} ~\sqrt{\frac{H_I}{v_{\star}}}.
\end{equation}
In the next subsection, we identify the regions in the landscape with maximal ${\cal VE_H}$  and show that they correspond to regions where there is a large hierarchy between the electroweak VEV and the cutoff, $\Lambda$.

\subsection{Maximisation of vacuum energy}

We now show that if the parameters of the potential are varied the vacuum energy contribution, ${\cal{VE_H}}$, from the Higgs sector is maximized for a small but finite electroweak  VEV. We show this in two steps. First, we consider the variation of $\mu_1^2$  and $\mu_2^2$,  keeping all the quartics fixed. This will give us some restrictions that must be imposed on the quartics for our mechanism to be successful.   In the next step, we will also vary the quartics and show how the regions allowed by the above conditions are automatically selected by our mechanism. As far as the parameters in \eq{vphi} are concerned, as already mentioned we must require the sign of $\mu^2_\phi$ to be fixed across the landscape. As far as the magnitude of, $\mu^2_\phi$ and $f$, are concerned we keep them also fixed in the following analysis for the sake of simplicity. As we will see later, however,  the impact of varying these two quantities can be understood in a straightforward way.

\subsubsection*{Step 1: Variation of $\mu_1^2$ and $\mu_2^2$}
To evaluate, ${\cal{VE_H}}$,  in different regions of the landscape, we classify the possible local minima of $V_H$ into three classes: (i)~minima with electroweak symmetry preserved (which necessarily requires $\langle \phi \rangle\neq 0$),  (ii)~minima with electroweak symmetry broken and $\langle \phi \rangle\neq 0$ and (iii)~minima with electroweak symmetry broken and $\langle \phi \rangle=0$.   We then evaluate the vacuum energy contribution for each case. We will see that the vacuum energy will peak for the last class of minima in regions of the landscape with a hierarchically small electroweak scale. Before carrying out such an analysis we must assume that the potential is bounded from below. This requires,  
\begin{equation}\label{norunaway}
\lambda_3+\lambda_4+ \hat{\lambda}_5+ 2\sqrt{\lambda_1 \lambda_2}\geq  0,
\end{equation}
for $\lambda_4+\hat{\lambda}_5<0$ and  $\lambda_3 \geq -2\sqrt{\lambda_1 \lambda_2}$ for $\lambda_4 + \hat{\lambda}_5 > 0$.

\paragraph{Class I -- Electroweak symmetry preserved :}
We first consider the case where $\langle H_1\rangle= \langle H_2 \rangle=0$ so that the trigger term is not effective. In these regions,   $\phi$ has a mexican hat potential that is minimized at, 
\begin{equation}
    \langle \phi \rangle= \pm\frac{f}{ \sqrt{4\lambda_\phi}}.
\end{equation}
${\cal VE_H}$  in this region only comes from   $V_\phi$ and is given by,
\begin{equation}
    {\cal VE_H}^I= -\frac{\mu_\phi^2 f^2}{16 \lambda_\phi}.
    \label{vbase}
\end{equation}
Note that these solutions correspond to minima only if both $\mu_1^2$ and $\mu_2^2$ are positive.

\paragraph{Class II -- EW symmetry broken and $\langle \phi \rangle \neq 0$ :} To analyze this possibility we  use the minimization condition with respect to the scalar $\phi$, i.e.  $\partial V_{H}/\partial \phi=0$, to obtain, 
\begin{eqnarray}
\hat{\phi}^2&=&\frac{f^2}{4 \lambda_\phi}-\frac{ \kappa f}{4 \lambda_\phi \mu_\phi}\ (H_1^\dagger H_2 +h.c.)
\end{eqnarray}
and then substitute this in the full potential, 
\begin{eqnarray}
V_H(\phi, H_1, H_2)|_{\phi\to \hat{\phi}}&=& -\frac{ \mu_\phi^2 f^2}{16 \lambda_\phi}+\hat{V}_{2HDM}(H_1, H_2).
\label{intout}
\end{eqnarray}
Here $\hat{V}_{2HDM}(H_1, H_2)$ is a 2HDM potential with the quartics in \eq{2hdmpot} modified,
\begin{eqnarray}
 \lambda_4 &\to& \lambda_4 - \frac{\kappa^2 }{8  \lambda_\phi}~~~~~~~~~~
\lambda_5 \to \lambda_5 - \frac{\kappa^2}{8\lambda_\phi}
\end{eqnarray}
 and an additional  term,
 \begin{equation}
\frac{\kappa \mu_\phi f}{8 \lambda_\phi}H_1^\dagger H_2+h.c.
\end{equation}
 We show in App.~\ref{2hdm} that if a minimum exists for $\hat{V}_{2HDM}$,  its contribution to the vacuum energy would be negative making the total vacuum energy in these local minima to be smaller than the value for class I minima given by \eq{vbase}.   On the other hand, $\hat{V}_{2HDM}$ has a runaway direction, and no EW breaking minima, if, 
\begin{equation}\label{nocoexist}
		\lambda_3 + \lambda_4 - \frac{\kappa^2}{8 \lambda_\phi} - \bigg\lvert \lambda_5 - \frac{\kappa^2}{8 \lambda_\phi} \bigg\rvert \leq - 2 \sqrt{\lambda_1 \lambda_2}.
	\end{equation}
and    $\lambda_4+\hat{\lambda}_5 < 0$. Class II minima are thus disallowed if the condition in \eq{nocoexist} is satisfied. Note that the left-hand side above is a monotonically decreasing function of $\kappa$ so that for a large enough   $\kappa$, \eq{nocoexist} this is guaranteed to be true.

 \paragraph{Class III -- Electroweak symmetry broken and $\langle \phi \rangle=0$ :} In regions of the landscape where $\langle H_1^\dagger H_2 \rangle \neq 0$, the trigger term   gives a new contribution to the $\phi^2$ term. This can  change the shape of the $\phi$-potential from a mexican hat to  one with a minimum at $\phi=0$, if,
\begin{equation}
			-\mu_\phi^2 +\frac{\kappa \mu_\phi}{f} \langle (H_1^\dagger H_2 +h.c.\rangle)\geq 0. 
\label{stability}
\end{equation}
  The scalar then gets stabilized at $\phi=0$ and the trigger term becomes ineffective giving no additional contribution to the 2HDM potential given by \eq{2hdmpot}. As we will soon show, this is the class of minima that will be selected by our mechanism.
  
  With $\langle \phi \rangle=0$, we must   minimise $V_{\text{2HDM}}$. As we discuss in App.~\ref{2hdm}, the VEVs of the doublets must take one of the following forms~\cite{diazcruz},

  	 \begin{eqnarray}
     	 \langle H_1 \rangle &=&\frac{1}{\sqrt{2}} \begin{pmatrix}
	 			0 \\
	 			v_1
	 		\end{pmatrix},\ \ \ \ \langle H_2 \rangle =\frac{1}{\sqrt{2}}  \begin{pmatrix}
	 			0 \\
	 			v_2 e^{i \xi}
	 		\end{pmatrix}; \label{vev1}\\
	 	\langle H_1 \rangle &= &\frac{1}{\sqrt{2}}\begin{pmatrix}
	 		0 \\
	 		v_1
	 	\end{pmatrix},\ \ \ \ \langle H_2 \rangle = \frac{1}{\sqrt{2}} \begin{pmatrix}
	 		u \\
			0
	 	\end{pmatrix}  \label{vev2}
	 \end{eqnarray}
	where $u,v_1, v_2>0$. The first possibility is realized if and only if,
 \begin{equation}
    \lambda_4+\hat{\lambda}_5<0,
    \label{photonmass}
 \end{equation}
whereas the second possibility is realized if  $\lambda_4+\hat{\lambda}_5>0$.\footnote{For $\lambda_4+\hat{\lambda}_5=0$ we get a flat direction in the EM breaking direction.}  Note, however, that in our set-up the second possibility is not consistent with \eq{stability} as it gives $\langle H_1^\dagger H_2\rangle=0$. This implies that for, $\lambda_4+\hat{\lambda}_5>0$, there are no stable class III minima.

 Assuming \eq{photonmass} holds so that the doublet VEVs are given by \eq{vev1}, we can now minimize with respect to $\xi$ to obtain, $\langle 2\xi+\alpha\rangle=0$ where $\alpha$ is the phase of $\lambda_5$ defined below \eq{2hdmpot} (see App.~\ref{2hdm}). Next, we can minimize the potential in the two CP-even, charge-neutral directions to obtain $v_1$ and $v_2$    as a function of $\mu_1^2, \mu_2^2$ and the quartics; these minimization conditions have been provided in \eq{muvev} of App.~\ref{2hdm}. Using these conditions we can trade $\mu_1^2$ and $\mu_2^2$ for $v_1$ and $v_2$---or equivalently for  $v=\sqrt{v_1^2+v_2^2}$ and $\tan \beta=v_2/v_1$ (see \eq{muvev2}).  
 
 As $V_\phi$ does not contribute to the vacuum energy,   the   vacuum energy contribution, ${\cal VE_H}$, for class III comes only from \eq{2hdmpot} and is given by (see App.~\ref{2hdm}), 
 \begin{equation}
     {\cal VE_H}^{III}=- \frac{1}{4}(\lambda_1 \, c_\beta^4 +\lambda_2 \, s_\beta^4+\lambda_{345} \, s_\beta^2 \, c_\beta^2) \, v^4,
     \label{vac2}
 \end{equation}
where $\lambda_{345} = \lambda_3 + \lambda_4 + \hat{\lambda}_5$, $s_\beta=\sin \beta$ and $c_\beta=\cos \beta$. The minimization conditions in \eq{muvev2} imply that as $\mu_1^2$ and $\mu_2^2$ vary in the landscape,   $v$ and $\tan \beta$ also vary. Let us now find the maximum value of  ${\cal VE_H}^{III}$  as $v$ and $\tan \beta$ are varied. It is clear that for a given value of $\beta$ the vacuum energy in \eq{vac2} monotonically decreases with increasing $v$. The maximal value of the vacuum energy is thus attained for the smallest   value of $v$ allowed by \eq{stability}, 
\begin{equation}
 v^2 \geq  \frac{\mu_\phi f}{\kappa s_\beta c_\beta}.
\end{equation}
 For a given value of $\tan \beta$ this gives the following upper bound for the vacuum energy for class III minima, 
\begin{equation}
     {\cal VE_H}^{III}\leq-\frac{1}{4} (\lambda_1 c_\beta^4 +\lambda_2 s_\beta^4+\lambda_{345} s_\beta^2 c_\beta^2) \frac{\mu_\phi^2 f^2}{\kappa^2  s^2_\beta c^2_\beta}.
 \end{equation}
Next, we maximize with respect to $\tan \beta$ to obtain the maximal value for the vacuum energy for class III minima, 
\begin{equation}
     {\cal VE_H}^{III, max}=-\frac{\mu^2_\phi f^2}{4 \kappa^2}(\lambda_{345} + 2 \sqrt{\lambda_1 \lambda_2})
     \label{ve3max}
 \end{equation}
 which is realized in the following region of the landscape,
\begin{equation}
    \label{position}
        v_\star^2 =\frac{\mu_\phi f}{\kappa s_{\beta \star} c_{\beta\star}},~~~~~	\tan^2 \beta_\star = \sqrt{\frac{\lambda_1}{\lambda_2}}
\end{equation}
which can also be written in terms of $\mu_{1\star}^2$ and $\mu_{2 \star}^2$---the value of the underlying parameters at this point---by inverting \eq{muvev2}.

Notice  that the value for the electroweak VEV in  \eq{position} can be much smaller than the cutoff $\Lambda \sim \sqrt{ H_I M_{pl}}$ (see \eq{cutoff}), 
\begin{equation}\label{hierarchy}
    v_\star^2 \ll \Lambda^2,
\end{equation}
as  $M\sim\sqrt{\mu_\phi f}$--- the global symmetry breaking scale associated with the PNGB $\phi$---can be naturally small. In fact, if we allow the magnitude of $\mu_\phi^2$ and $f$ to vary in the landscape while keeping the sign of the former fixed, the vacua with the smallest value of $\mu_\phi f$ will be selected as they will correspond to the highest value of ${\cal VE_H}^{III, max}$.

We will now show that the region of the landscape defined by \eq{position} can have the highest vacuum energy contribution,  ${\cal VE_H}$, not only among class~III minima but in the whole landscape, provided certain conditions are satisfied by the quartics. The desired region given by \eq{position} will then be selected by our mechanism. The first of these conditions arises from requiring  that     ${\cal VE}_{H}^{III,max}$   exceeds the vacuum energy for class~I minima, ${\cal VE}_H^I$ which gives,
	\begin{equation}
		\kappa^2 > {4}\lambda_\phi (\lambda_{345} + 2 \sqrt{\lambda_1 \lambda_2}) \; .
        \label{bigger}
	\end{equation}
We will demand one more condition on the quartics to ensure that \eq{position} is the region with maximal vacuum energy in the whole landscape. We want to eliminate the possibility of the coexistence of multiple local minima for a given choice of the parameters, $\{\alpha_H^i\}$. This is important because, for more than one minimum, the vacuum energy computation would involve solving the FPV equation to obtain the relative probability of the fields to be in the different minima; in particular \eq{ve3max} is not valid if there are other coexisting minima. We show in App.~\ref{2hdm} that for a given choice of parameters, $\{\alpha_H^i\}$,  class III minima cannot coexist with class~I minima. On the other hand, class II and III minima may coexist.  If the condition in \eq{nocoexist} is imposed, however, stable class II minima are disallowed. In this case, only class I and III  minima can exist---and if \eq{bigger} is satisfied---the class III minima satisfying \eq{position} will have the maximal vacuum energy in the whole of the landscape. 
     
\begin{figure}[t!]
		\centering
		\includegraphics[width=1\linewidth]{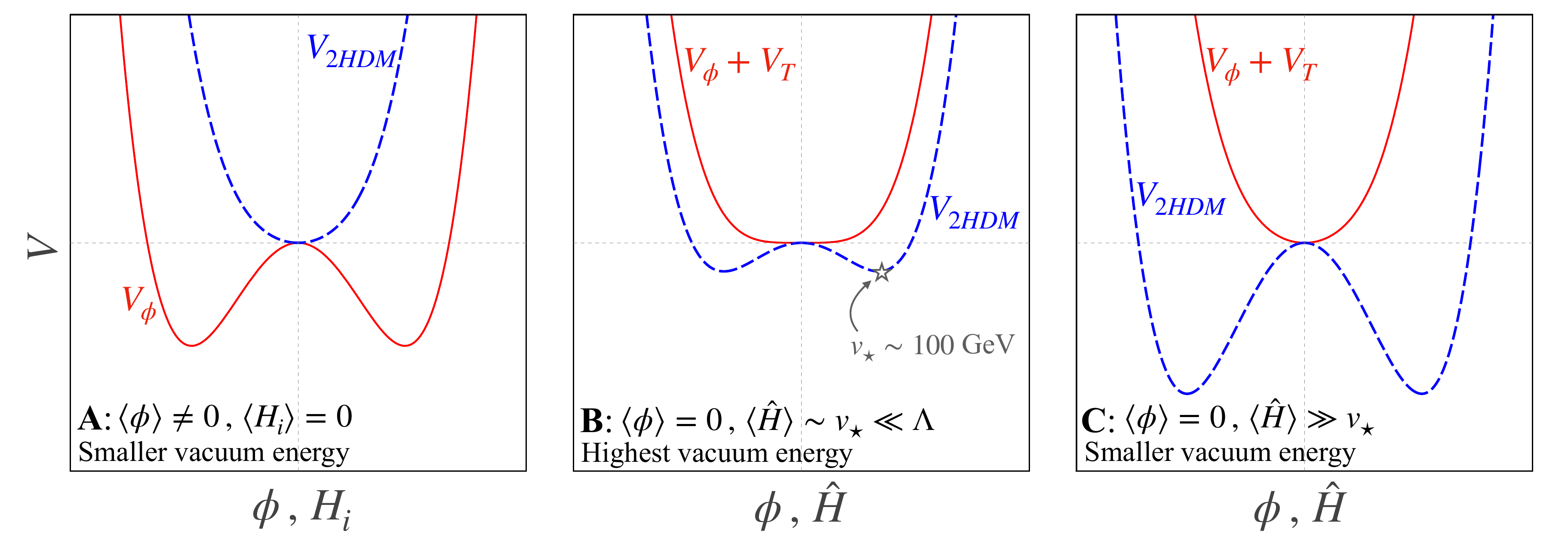}
		\caption{The potential of $\phi$ and the Higgs doublets for varying values of the electroweak VEV, $v$. In the left panel, we show a class I minimum with $v=0$ and a mexican hat potential for $\phi$. $H_i$ in this case can be any of the 8 components of the 2 Higgs doublets. In the middle panel, we take, $v=v_\star$, the threshold value required to change the $\phi$-potential to one with its minimum at the origin. In the right panel we show that for, $v\gg v_\star$, while the $\phi$-potential is still minimized at the origin, the Higgs potential becomes deeper. In the middle and right panel, the minima are of the class III category and $\hat{H}=\sqrt{\phi_3^2+\phi_7^2}$, where $\phi_3$ and $\phi_7$ are the two CP even components (see \eq{components}). The  middle panel clearly corresponds to vacua with maximal vacuum energy. The three scenarios have been labeled A, B, and C as they directly correspond to the three similarly labeled categories of vacua in Fig.~\ref{fig:1}.}
		\label{fig:2}
	\end{figure}

To sum up, we have shown that the region of the landscape defined by \eq{position} has the maximal vacuum energy in the whole landscape if the  following conditions are met:
\begin{itemize}
\item{the potential is bounded from below which requires \eq{norunaway},   i.e., \eq{norunaway} is satisfied}  
\item{class III minima  exist, which requires \eq{photonmass},}   
\item{the maximal vacuum energy of the class~III minima exceeds that of class~I minima,  ${\cal VE_H}^{III, max}>{\cal VE_H}^I$ which gives \eq{bigger}, and,}    \item{class II minima do not exist which is implied by \eq{nocoexist}.}    
\end{itemize}
 We illustrate the basic idea of our mechanism in Fig.~\ref{fig:2} where we show the shape of the scalar potentials as the electroweak VEV is varied. We show only class I and III minima in Fig.~\ref{fig:2} assuming \eq{nocoexist} is obeyed.  Given  \eq{nocoexist}, a positive $\mu_1^2$ and $\mu_2^2$ results in class I minima whereas even if one of these two parameters is negative only class III minima can be obtained.  In the panel on the left, we show the situation when the electroweak VEV is zero and the pseudoscalar has a mexican hat potential. In the middle panel, the electroweak VEV is at the threshold value given by \eq{position} such that the $\phi$-potential is minimized at the origin. If the vacuum energy in the middle panel has to be larger than the one in the left panel, the gain in vacuum energy due to the rising of the $\phi$ minima must be larger than the negative contribution from the electroweak VEV; it is precisely this requirement that yields \eq{bigger}. Finally, on the right, we show that as the electroweak VEV increases the Higgs minimum becomes deeper resulting in a monotonically decreasing vacuum energy contribution as in the SM. Thus the maximal value of the vacuum energy contribution, ${\cal VE_H}$, is attained in the middle panel where the electroweak VEV is much smaller than the cutoff and the $\phi$ resides at the origin.   This region then expands exponentially faster than other regions of the landscape to eventually occupy almost all of the multiverse---thus resolving the hierarchy problem.

\subsubsection*{Step 2: Variation of the  quartics}

We will now show that if the quartics, $\lambda_{1-5}$ and $\kappa$ are varied, we do not need to impose the conditions on the quartics listed in step 1, as the regions of the landscape satisfying them would be automatically selected by our mechanism.  First, regions of the multiverse where \eq{norunaway} is not obeyed have a runaway direction in field space which would result in a big crunch. Next, in regions of the landscape where,   \eq{photonmass} or \eq{bigger}, is not satisfied either class III minima do not exist at all or they have vacuum energy less than that of class I minima.

In regions where the last condition, i.e.  \eq{nocoexist} is not satisfied, class II minima become possible. As we have already mentioned, and discussed in detail in App.~\ref{2hdm}, such minima will always have vacuum energy smaller than ${\cal VE_H}^{I}$ and, if \eq{bigger} is satisfied, also smaller than ${\cal VE_H}^{III, max}$. It is possible, however, for a class II minimum to coexist with a class III minimum having higher vacuum energy. We would then need to obtain $P(\phi, H_1, H_2)$ from the FPV equation to compute the vacuum energy.  An upper bound on the vacuum energy in this scenario can be obtained by considering a $P(\phi, H_1, H_2)$ that peaks entirely on the higher class III minimum; we can then use the expression for   ${\cal VE_H}^{III, max}$ in \eq{ve3max} to compute this upper bound. The value of ${\cal VE_H}^{III,max}$, however, would be higher in regions where  \eq{nocoexist} is respected compared to regions where this condition is violated.  The reason for this is that the left hand side of \eq{nocoexist} is a monotonically decreasing function of $\kappa$ so that regions obeying this condition will have a higher  $\kappa$  and thus a higher, ${\cal VE_H}^{III,max}$ (see \eq{ve3max}).

Thus the regions of the landscape satisfying the conditions, \eq{norunaway}, \eq{photonmass}, \eq{bigger}, and \eq{nocoexist}, have larger vacuum energy than regions violating them. Our mechanism, therefore, automatically selects them.

\subsection{Volume of desired region}
\label{measure}
We have established that in our model the maximum vacuum energy is attained in parts of the landscape where the electroweak VEV is finite but much smaller than the cutoff.  We now wish to quantitatively show that, after a sufficiently long time, these regions with maximal vacuum energy given by,  ${\cal VE_H}^{III, max}$ in \eq{ve3max}, dominate the multiverse in volume over other regions. 

Following Ref.~\cite{linde1}, we first show that the universe is dominated by the maximal energy vacuum state in the landscape, that we call  $\lambda_*=\{i_\star, j_\star, k_\star, m_\star\}$, where $i_\star$ and  $j_\star$  respectively specify the choice of parameters in the Higgs sector and the inflaton sector;  $k_\star$ denotes the choice of the cosmological constant; and $m_*$ labels the highest vacuum of the inflaton, $\chi$ (see \eq{vace}).  The volume weighted probability distribution functions for the different vacua in the landscape    obey the following differential equations (see for eg.  Ref.~\cite{linde1} and the references therein), 
\begin{eqnarray}
    \frac{d P_{\star}}{dt}&=&-\sum_\lambda^{\lambda\neq\lambda_{\star}} \Gamma_{{\lambda_\star} \to \lambda} P_{\star}+ \sum_{\lambda}^{\lambda\neq\lambda_{\star}} \lambda \Gamma_{{\lambda } \to \lambda_\star} P_{\lambda}+ 3 H_\star P_\star \label{vacua1}\\
    \frac{d P_{\lambda}}{dt}&=&-\sum_{\mu}^{\mu \neq  \lambda} \Gamma_{\lambda \to \mu} P_{\lambda} +\sum_{\mu}^{\mu \neq \lambda,\lambda_*} \Gamma_{\mu \to \lambda} P_\mu + \Gamma_{{\lambda_\star} \to \lambda} P_{{\star}}+  3 H_\lambda P_\lambda \label{vacua2}
\end{eqnarray}
where $P_{\lambda}$ is the volume-weighted probability to be in the state $\lambda$, $P_\star=P_{\lambda_\star}$ is the volume-weighted probability to be in the maximal energy vacuum and $H_\star$ is the corresponding Hubble scale.  The transition rates from one vacuum to another, $\Gamma_{\alpha \to \beta}$,  are, as we will soon discuss, (doubly) exponentially suppressed numbers. Thus we can take $H_\star$ to be larger than all other parameters appearing in the above equations. This results in $P_\star$ dominating over the other $P_\lambda$ so that, the last term in the right hand side (RHS) of \eq{vacua1} and the third term in the RHS of  \eq{vacua2}, become the dominant terms. The solution to these equations is found to have the following  form for large times,
\begin{equation}
    P_\star(t)= P_\star(0) e^{3 H_\star t}, \qquad P_\lambda (t)= \frac{\Gamma_{\lambda_\star \to \lambda }}{H_\star} P_\star(t). 
\end{equation}
As $ {\Gamma_{\lambda_\star \to \lambda }}/{H_\star} \ll 1$, we see that  the maximal energy vacuum,  $\lambda_*$, dominates over all others. The inflationary Hubble scale, $H_I$, in Sec.~\ref{land} should thus be identified with the Hubble scale in this vacuum, i.e. $H_I= H_\star$. The maximal energy vacuum, $\lambda_\star$, is the one for which all the terms in \eq{vace}, including in particular the Higgs-sector contribution,  ${\cal VE}_H$, are independently maximized.\footnote{Here we are assuming that for the maximal value of the cosmological constant, $\Lambda_{cc}^{(k_*)}$ the full range of variation in the parameters, $\{\alpha_H^{(i)}\}$, and,  $\{\beta_\chi^{(j)}\}$ in \eq{vace}, is realized in the landscape.} In this vacuum, the electroweak VEV is thus given by \eq{position} and is much smaller than the cutoff. We see that all the, $P_\lambda$, eventually enter a stationary regime where they grow at the same rate, $\ e^{3 H_\star t}$. This is because regions in a vacuum state, $\lambda \neq \lambda_\star$,  primarily arise due to tunneling from the maximal energy vacuum.

In order to give rise to a universe such as ours the inflaton sector must transition to vacua, $\lambda_\star \to \lambda_{slow}$, which permit a slow-roll and reheating phase. The total volume of such universes  are  given by,
 \begin{eqnarray}\label{vt}
{\cal V}(t)&=&\sum _{\lambda_{slow}}P_\star(0) e^{3 H_{\star} (t-t_+)} \Gamma_{\lambda_\star \to \lambda_{slow}} e^{N_{slow}}r_V(t_{age})\Theta(t-t_+)
\end{eqnarray}
where $t_{age}$  is the age of the universe since reheating and $r_V(t_{age})$ is the factor by which the universe has expanded in this time period;  $N_{slow}$ and $t_{slow}$ are, respectively,  the number of e-folds and the time elapsed during the slow-roll phase in the vacuum $\lambda_{slow}$;  $\Gamma_{\lambda_\star \to \lambda_{slow}}$ is the rate for the transition, $\lambda_\star \to \lambda_{slow}$, and   $t_+=t_{age}+t_{slow}$.  The theta function arises because a minimum time, $t_+$, is required before any universe of an age, $t_{age}$, can emerge.  Our assumption of complete decoupling between the Higgs and inflaton sector (see Sec.~\ref{land}) becomes crucial here and implies that the transition, $ {\lambda_\star \to \lambda_{slow}}$---or more explicitly $\{i_\star, j_\star, k_\star, m_\star\} \to \{i_\star, j_{slow}, k_{slow}, m_{slow}\}$---does not result in a change in the Higgs sector parameters, $\{\alpha_{H}^{i}\}$, and the electroweak VEV  remains, $v_\star^2 \ll \Lambda^2$, given by \eq{position}. Note that  number of e-folds in   slow-roll phase   need not be exponentially large in our model.

 We now compute the volume of the regions of the multiverse that undergo slow-roll and reheating but with $v\neq v_\star$. Using again our assumption of decoupling between the Higgs and inflaton sectors, we see that these regions will emerge from the maximal energy vacuum $\lambda_*$ after two transitions, i.e. $ {\lambda_\star \to \lambda' \to \lambda'_{slow}}$---or written explicitly $\{i_\star, j_\star, k_\star, m_\star\} \to \{i', j_{\star}, k', m_{\star}\} \to \{i', j_{slow}, k'_{slow}, m_{slow}\}$---where the first one changes Higgs sector parameters and the cosmological constant but not the inflaton sector parameters. The second transition, on the other hand,  changes the inflaton sector parameters and the cosmological constant but not the Higgs sector parameters. Note that the inflaton sector in the vacuum states, $\lambda_{\star}$ and $\lambda'$, are identical up to a difference in the cosmological constant, $\Delta \Lambda_{cc}$. This is also the case for the vacua, $\lambda_{sol}$ (in \eq{vt}), and $\lambda'_{sol}$ and the difference in cosmological constants  is, in fact,   identical to the previous case, i.e. $\Delta \Lambda_{cc}$. This implies that the transition rates,   $\Gamma_{\lambda_\star \to \lambda_{slow}}$ and $\Gamma_{\lambda' \to \lambda'_{slow}}$,  are equal. For the volume arising after the first transition, we get,
 \begin{equation}\label{doublexp}
P'(t)=\sum_{\lambda'} P_\lambda (t)=\sum _{\lambda'} \frac{\Gamma_{\lambda_\star \to \lambda' }}{H_\star} P_\star(t) .
 \end{equation}
  Note that this volume is exponentially smaller than, $P_\star(t)$, despite the fact that the summation above may run over an exponentially large number of states (for a system of $n$ scalars the number of vacua scales as $e^n$). This is because the tunneling rate $\Gamma_{\lambda_\star \to \lambda' }$ is suppressed by the double exponential, $e^{-S}$, where the action $S$ is itself an exponentially large ratio of energy scales raised to the fourth power.\footnote{We thank S. Trivedi for explaining this point to us.} Even in the string theory landscape, while the number of vacua has been estimated to be as large as $10^{500}$~\cite{Douglas}, the estimate for the probabilities, $e^{-10^{122}}$~\cite{kklt1}, is exponentially smaller. 
  
  We can again compute the volume of the universes arising from $P'(t)$ that undergo a slow-roll and reheating phase by considering the second transition, $\lambda' \to \lambda'_{slow}$,
  \begin{eqnarray}\label{vpt}
{\cal V}'(t)&=&\sum_{\lambda'_{slow}}P'(t-t_+)   \Gamma_{\lambda_\star \to \lambda_{slow}} e^{N'_{slow}}r_V(t_{age})\Theta(t-t'_+),
\end{eqnarray}
where we have used, $\Gamma_{\lambda' \to \lambda'_{slow}}=\Gamma_{\lambda_\star \to \lambda_{slow}}$. Here,  $N'_{slow}$ and $t'_{slow}$, the number of e-folds and slow-roll time in the vacuum state, $\lambda'_{slow}$, are smaller than the corresponding quantities, $N_{slow}$ and $t_{slow}$, in \eq{vt}. To show this, we first note that the vacuum energy of, $\lambda'_{slow}$, is smaller than the vacuum energy of  $\lambda_{slow}$. This follows from the fact that although the change in vacuum energy for the $\lambda_\star \to \lambda_{slow}$ transition is the same as that in the  $\lambda' \to \lambda'_{slow}$ transition,    the vacuum $\lambda_\star$ has greater vacuum energy than $\lambda'$ by definition.  Thus, despite the inflaton potential being the same in both cases, the Hubble parameter in $\lambda'_{slow}$ is smaller than that in $\lambda_{slow}$ resulting in a shorter slow roll time and smaller number of e-folds.

We must address a final subtlety before quantifying the volume in the multiverse with a large electroweak hierarchy.  We want to compare this volume with the volume without such a hierarchy for some large time, $t$. There is, however,  no well-defined way of choosing time slices across the causally disconnected patches in the multiverse which leads to the so-called measure problem. In the calculation of ${\cal V}(t)$ and ${\cal V}'(t))$ we have been implicitly using the proper time cutoff measure where the time slices are chosen according to the proper time elapsed in each patch. This measure however leads to the so-called youngness paradox~\cite{guthYP}. This can be understood directly from, \eq{vt}, if we compare volumes for universes with different values of $t_{age}$. We see that a smaller value of $t_{age}$ yields a smaller $t_+$ and thus an exponentially larger volume. Thus younger universes are exponentially favoured in the proper time cutoff measure which results in this paradox. The paradox can be restated as an exponential preference for a higher than observed CMB temperature~\cite{tegmark} which observationally rules out the proper time cutoff measure.

 Here we will use the stationary measure~\cite{linde1} which evades the youngness paradox as well as other issues such as a gauge dependence on the time parameterization and the Boltzmann brain problem~\cite{linde2, linde3}. In the stationary measure we shift the time coordinate such that volumes are evaluated only as a function of the time elapsed since the beginning of the stationary regime. In our case, we obtain for  the two volumes in \eq{vt} and \eq{vpt} in  the stationary measure, 
\begin{eqnarray}\label{vol}
{\cal V}(t)_{stationary}&=&{\cal V}(t)\vert_{t\to t+t_+} \nonumber\\
{\cal V}'(t)_{stationary}&=&{\cal V}'(t)\vert_{t\to t+t'_+} 
\end{eqnarray}
This removes the dependence of ${\cal V}(t)_{stationary}$ on $t_+$ and thus there is no youngness paradox in the stationary measure. We can now evaluate the ratio of the latter volume to the former for large values of $t$, 
\begin{equation}\label{main}
    \lim_{t\to \infty}\frac{{\cal V}'(t)_{stationary}}{{\cal V}(t)_{stationary}}\ll 1
\end{equation}
where we have used $N'_{slow}<N_{slow}$ and the arguments below \eq{doublexp}.\footnote{Note that we have not addressed the cosmological constant problem in our discussion so far. We would like to emphasize that our mechanism is completely compatible with Weinberg's anthropic solution to the cosmological constant problem~\cite{Weinberg}. We can ensure that the final cosmological constant is in the anthropic range by multiplying both ${\cal V}(t)_{stationary}$ and ${\cal V}'(t)_{stationary}$ by anthropic suppression factors. Assuming these factors are of the same order, our conclusion in \eq{main} would remain unchanged.} This shows that in the stationary measure, after a sufficiently long time,  the volume of the multiverse is dominated by regions where the electroweak scale is given by \eq{position} and is thus hierarchically smaller than the cutoff, $\Lambda$.

\section{Phenomenology of the 2HDM sector}
\label{sec:2hdm}
The phenomenology of the 2HDM sector provides an opportunity to test the most important prediction of our model, namely that of the value of $\tan \beta$, 
\begin{equation}\label{pred}
		\text{tan}^4\beta = \left(\frac{\lambda_1}{\lambda_2}\right),
	\end{equation}
derived  in  \eq{position}. Indeed, using the standard expressions for the masses and mixing angles of the different Higgs bosons (see for eg. Ref.~\cite{Arbey:2017gmh})  one can convert the above equation into the following  prediction, 
\begin{equation}\label{prediction}
\text{cos}~\alpha =  \sqrt{\frac{m_h^2- m_H^2 \text{tan}^2\beta}{\left(m_h^2- m_H^2\right)\left(1 + \text{tan}^2\beta\right)}}.
\end{equation}
that connects, $\alpha$, the mixing angle between the CP-even Higgs bosons to  $\tan \beta$ and $m_H$, the mass of the heavier CP even Higgs.  These are all observables that can be independently measured. While $m_H$ can be measured by directly producing $H$ at LHC, the measurement of the couplings of the charged Higgs, $H^+$, and pseudoscalar, $A$, can determine $\tan \beta$. For a given value of $\tan \beta$, a measurement of the couplings of the SM-like Higgs, $h$, can then determine  $\alpha$.

   \begin{figure}[t!]
		\centering
		\includegraphics[width = 0.8\linewidth]{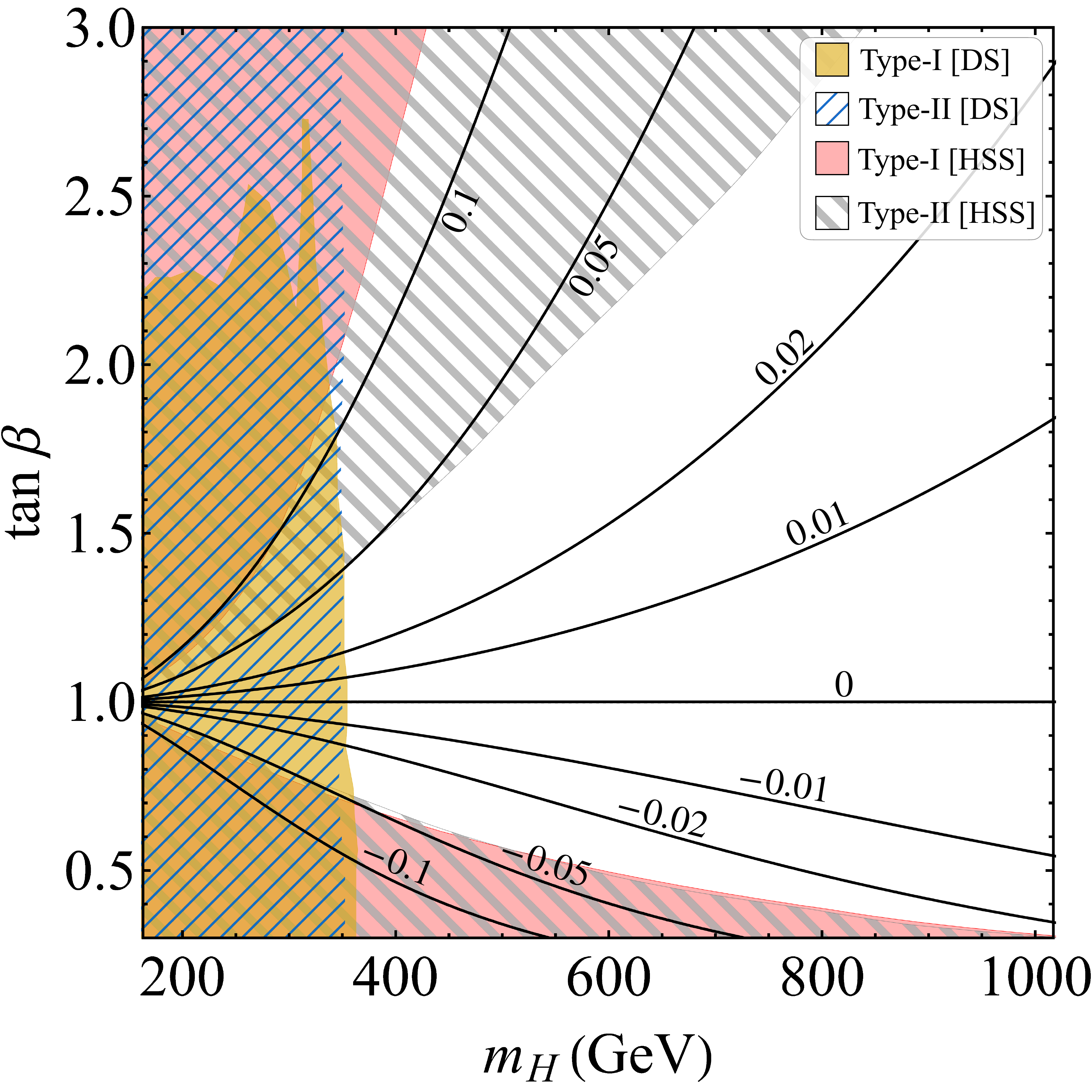}
		\caption{The 2HDM parameter space for our model in the $\text{tan\,}\beta\,$-$\,m_H$ plane. The black contours show predictions for  $\cos (\beta-\alpha)$ as a function of  $m_H$ and $\tan \beta$. These predictions can be verified by measurement of the couplings of the SM-like Higgs whose deviations from SM are controlled by $\text{cos}(\beta-\alpha)$. We show the values of  $\text{cos}(\beta-\alpha)$    excluded by Higgs signal strength (HSS) measurements for both Type I (pink) and Type II models (gray hatched). We also show direct search (DS) bounds for  Type-1 (blue hatched)  and Type-2 (brown) models (see text for details).}
		\label{fig:4}
	\end{figure}

 We show this graphically in Fig.~\ref{fig:4} where contours of $\cos(\beta-\alpha)$ predicted by \eq{pred} have been shown on the $\tan \beta$-$m_H$ plane.  Note that, $\cos(\beta-\alpha)$, is crucial in determining the deviations of the couplings of the lighter  Higgs, $h$, from SM values. In particular $\cos(\beta-\alpha)=0$  is the alignment limit where couplings of the $h$ exactly match SM values. Too-large values of $\cos(\beta-\alpha)$   are thus excluded by  Higgs signal strength measurements; Fig.~\ref{fig:4} shows the resulting bounds obtained by Ref.~\cite{lenz1, lenz2}. Bounds for Type I models where only $H_1$ couples to all the fermions have been shown separately from bounds on Type II models where $H_2$ couples to up-type fermions and $H_1$ couples to down type fermions and leptons. We have shown only ${\cal O}(1)$ values of $\tan \beta$ motivated by \eq{pred} as much larger or smaller values will require a large hierarchy in the values of $\lambda_1$ and $\lambda_2$. 
 
 In Fig.~\ref{fig:4} we also show the direct search bounds on the 2HDM parameter space.  The hatched region shows the direct search bounds on  $H$ and $A$ presented in Ref.~\cite{Arbey:2017gmh}, These arise from the processes $H \to \gamma \gamma$ and $H/A \to \tau \tau$.  For the latter process, the pseudoscalar and heavy CP even Higgs masses have been assumed to be equal,  i.e.  $m_A=m_H$.  This assumption results in conservative bounds as  $m_A$ can be independently varied in our model and a larger $m_A$  which would result in weaker bounds.  As far as the charged Higgs is concerned, the most stringent bounds arise from  $b \to s$ processes~\cite{Arbey:2017gmh}. For a given $\tan \beta$, these can be always satisfied by choosing a large enough charged Higgs mass $m_{H^+}$. While this requires somewhat large absolute values of the coupling $|\lambda_{45}|=|\lambda_4+\lambda_5|$ in certain regions of the parameter space---for instance, we require $|\lambda_{45}|>15$ for $\tan \beta < 0.8 $ ($\tan \beta < 1$)  in Type II (Type I) models---this is still well within the perturbativity bound on the quartics, $\lambda_i\lesssim 16 \pi^2$.

\section{Phenomenology of $\phi$ and dark matter}
\label{phipheno}

The pseudoscalar, $\phi$, in our model, is a light weakly coupled state with its mass and couplings suppressed by powers of $f$. As the $\phi \to -\phi$ symmetry remains unbroken in our selected vacua, the $\phi$ couples quadratically to SM fields. These quadratic couplings, 
\begin{eqnarray}\label{lagphi}
{\cal L}_{\phi^2 SM}=-c_{e}\frac{\phi^2}{f^2}m_e \bar{e}e-c_{q}\frac{\phi^2}{f^2}m_q \bar{q}q+c_{\gamma}\frac{\phi^2}{f^2}\frac{\alpha_{em}}{4 \pi} F_{\mu \nu}F^{\mu \nu}+c_{g}\frac{\phi^2}{f^2}\frac{\alpha_{s}}{4 \pi} G^a_{\mu \nu}G^{a,\mu \nu}
\end{eqnarray}
 can, for instance, be generated via a tree-level Higgs exchange due to the trigger term in \eq{trigger}. The observational and experimental bounds on such quadratically coupled light fields are negligible unless there is a background classical configuration of $\phi$ (see for instance Ref.~\cite{gilad}). We will thus study how a classical background configuration of $\phi$ can arise due to the misalignment mechanism. We will also show that in certain regions of the parameter space, $\phi$ can account for the observed dark matter relic density.

As far as the mass of $\phi$ is concerned, an additional subtlety is introduced by the fact that, in our model, it is actually smaller than the naive expectation, ${\cal O}(\mu_\phi^2)$. This is because the mass of $\phi$ is given by, 
\begin{equation}\label{cancel}
    m_\phi^2=\left(-\mu_\phi^2+\kappa \frac{\mu_\phi}{f}v_\star^2 s_{\beta_\star} c_{\beta_\star}\right),
\end{equation}
where a cancellation takes place between the two terms above in the selected vacua (see \eq{position}). The cancellation between the two terms on the right-hand side, however, is only approximate as $\kappa$, and the 2HDM parameters are scanned in the landscape in discrete steps so that the second term above also takes a discrete set of values.  To take into account this fact we take $m_\phi^2= \epsilon^2 \mu_\phi^2$, where,    $\epsilon \ll 1$,  is a parameter that captures the degree of cancellation between the two terms---which in turn depends on how finely the underlying Higgs sector parameters are being scanned. In particular,  the limit of continuous scanning of parameters corresponds to  $\epsilon \to 0$.

To study the cosmology of $\phi$ after reheating we  thus consider the following leading terms in its potential 
\begin{equation}\label{potphi}
    V_\phi=  m_\phi^2 \phi^2 +\lambda_\phi \frac{\mu_\phi^2}{f^2}{\phi^4}
\end{equation}
and treat $\epsilon=m_\phi/\mu_\phi$ as a free parameter. The initial misalignment of the field ultimately arises from the quantum spreading of the field, $\phi$, during inflation. If the reheating temperature, $T_{RH}$, is smaller than the weak scale, i.e. $T_{RH}\lesssim v \sim \sqrt{\mu_\phi f}$, we do not expect any significant thermal corrections to either the 2HDM or the  $\phi$-potential. At higher temperatures, thermal corrections to the $\phi$-potential are model-dependent and depend on the UV origin of the $\phi$-field. For the sake of definiteness, we assume here that $\phi$ is the axion of a strongly coupled sector with a confinement scale given by ${M}\sim\sqrt{\mu_\phi f} \sim v$.  The potential in \eq{potphi} will thus vanish for, $T_{RH} \gtrsim v$, and reappear only when the temperature drops below the weak scale. As the $\phi$-potential becomes flat for such high reheating temperatures, we assume that $\phi$ retains its position in field space during this process. We thus assume that the order of the initial misalignment, $\delta\phi_m$,  is given by the amount of quantum spreading of $\phi$ during inflation.

To study the post-inflationary cosmology of $\phi$ we must keep in mind two important scales, one of them being the initial misalignment scale, $\delta\phi_m$.  The other scale corresponds to  the  $\phi$-field value up to which the quadratic term dominates over the quartic one, 
\begin{equation}
    \delta \phi_q\sim \epsilon f/\sqrt{\lambda_\phi}.
    \label{phipot}
\end{equation}
The cosmology of $\phi$ depends on the relative value of these two scales. If $\delta\phi_m \gg \delta\phi_q$, we can ignore the small region where the quadratic term becomes relevant, whereas if, $\delta\phi_m \ll \delta\phi_q$,  $\phi$ only feels the quadratic part of the potential so that we can effectively ignore the quartic and other higher order terms. In App.~\ref{FPV} we estimate the quantum spreading of $\phi$ during inflation, and thus  the misalignment scale to be, 
\begin{eqnarray}\label{misalign}
    \delta \phi_m &\sim&  \frac{H_I^2}{m_\phi}~~~~~~~~~~~~~~~~~~~~~(\delta \phi_m \ll \delta \phi_q) \nonumber\\
    \delta \phi_m &\sim&  \frac{H_I}{\lambda_\phi^{1/4}}\sqrt{\frac{f}{\mu_\phi}}~~~~~~~~~~~~~(\delta \phi_m \gg \delta \phi_q).
\end{eqnarray}
Using either of the two expressions for $\delta \phi_m$, we can recast the condition to be in the quadratic  regime   as follows,  \begin{eqnarray}\label{req}
H_I\lesssim \epsilon \sqrt{\mu_\phi f}/\lambda_\phi^{1/4},
\end{eqnarray} 
where the reverse inequality is required for the quartic regime. We now consider both these regimes.

  \subsection{Quartic regime}
  
 If the Higgs sector parameters are scanned very finely we are in the limit, $\epsilon\ll 1$, so that we can take  $\delta\phi_m \gg \delta\phi_q$ and effectively ignore the quadratic term in \eq{phipot}. To study the cosmological evolution of, $\phi$, after reheating we need to solve its equation of motion,
 \begin{equation}\label{eom}
\ddot{\phi} + 3 H(T) \dot{\phi} +V'(\phi)=0
\end{equation}
  where, $H$, is the Hubble scale, $T$ is the temperature  and we take $V(\phi)=\lambda_\phi \frac{\mu_\phi^2}{f^2}{\phi^4}$. This equation has been studied in detail in Ref.~\cite{Turner, Masso}. The above equation admits an  oscillatory solution for  $H(T)\lesssim \nu$, where $\nu$ is the frequency given by, 
\begin{equation}
\nu= \frac{4~\Gamma\left(\frac{3}{4}\right)}{2 \sqrt{2 \pi} \Gamma \left(\frac{1}{4}\right)} \left(\frac{\lambda_\phi\mu_\phi^2}{f^2} \rho_\phi\right)^{1/4} 
\end{equation}
 and $\rho_\phi$ is the energy density of the $\phi$-field. 
 
 The oscillations begin at a temperature, $T_{osc}= \min\{v, T_{RH}, T_{sol}\}$, $T_{sol}$, being the temperature for which, $H(T_{sol})= \nu$. Once the oscillatory behavior begins, the energy density in the $\phi$-field scales as $~1/a^4$. Thus the $\phi$ effectively acts like a source of dark radiation that is constrained to be much smaller than the energy density in the SM sector.  If, at the beginning of the oscillations, we require the energy density of $\phi$ to be much smaller than the expected energy density from standard cosmology,  it will remain so throughout cosmological history.\footnote{Here we are assuming that $\epsilon$ is small enough such that  the $\phi$-field value never becomes smaller than $\delta \phi_q$ due to the $1/a$  damping. } We find that if $T_{osc}= T_{sol}$ or $T_{osc}=v$  this requirement is automatically satisfied, where we have used  \eq{cbeatsq} and the fact that the initial value of $\rho_\phi=\lambda_\phi \frac{\mu_\phi^2}{f^2}(\delta \phi_m)^4\sim H_I^4$ (see \eq{misalign}). If on the other hand, $T_{osc}=v$, we must require $H_I^4 \ll g_\star(T_{RH}) T^4_{RH}$, $g_\star(T_{RH})$ being the number of relativistic degrees of freedom at the reheating temperature.

  \subsection{Quadratic regime and wave-like dark matter}

In the quadratic regime, we take $V_\phi=m_\phi^2 \phi^2/2$ in \eq{eom}. As is well known, this results in damped oscillations of $\phi$ that can account for dark matter. The  dark matter abundance is given by (see for instance~\cite{kolb}), 
\begin{equation}
    \Omega_\phi h^2 \approx 3 \, (\Delta \theta_m)^2 \bigg( \frac{m_\phi}{ {\text{ 1~eV}}}\bigg)^2 \bigg(\frac{f}{10^{9} {\text{ GeV}}}\bigg)^2 \bigg(\frac{100 {\text{ GeV}}}{T_{osc}}\bigg)^3 \; .
\end{equation}
where, $T_{osc} \sim \text{min} \Big[v, T_{RH}, T_{sol} \Big] $, is  the temperature at which the oscillations begin. Here  $T_{sol}$ is determined by solving $3 H (T_{sol}) = m_\phi$.

\begin{figure}[h!]
		\centering
		\includegraphics[width = 0.8\linewidth]{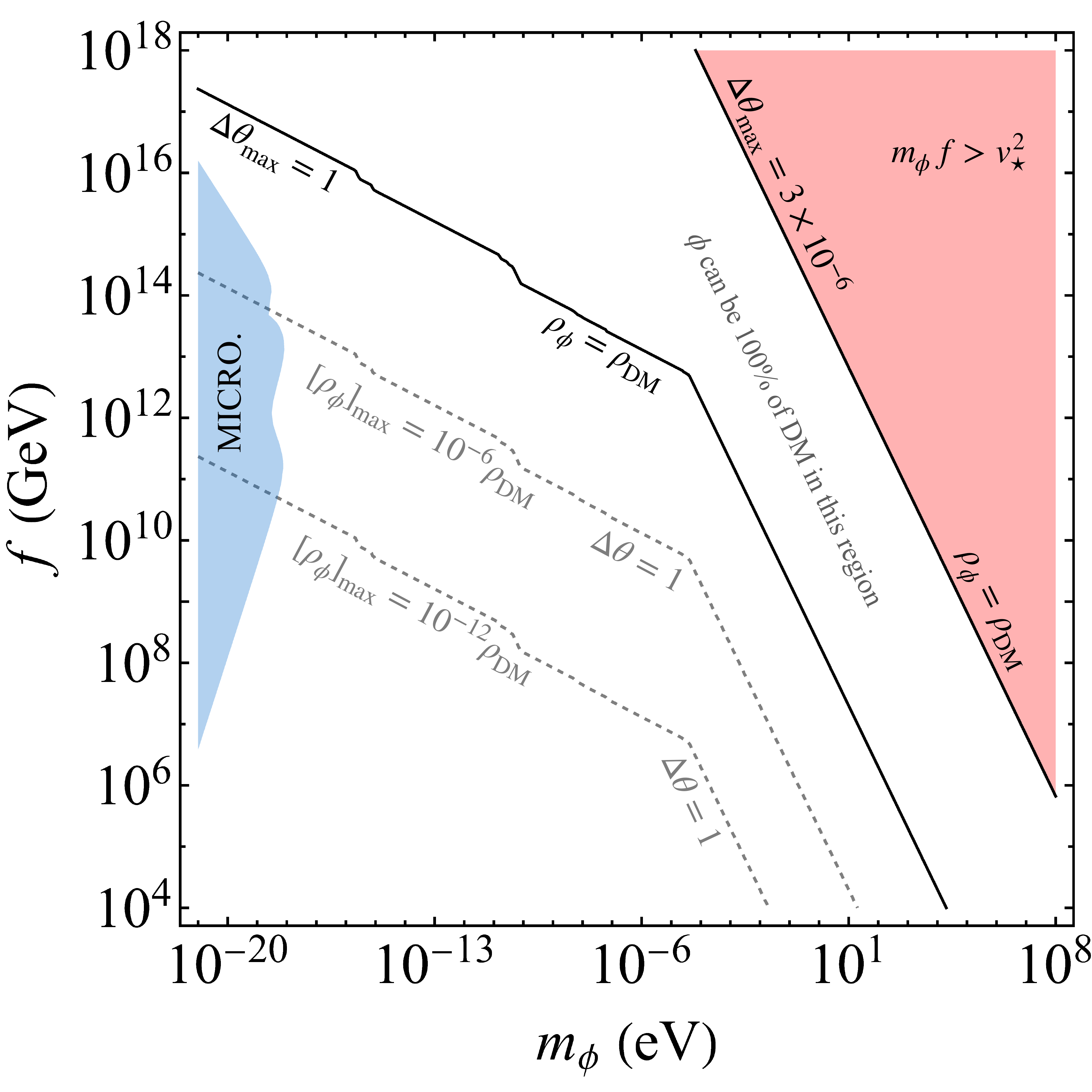}
		\caption{The region in the $m_\phi$-$f$  parameter space where the pseudoscalar, $\phi$, can account for the observed dark matter of the universe. In the region between the two solid black lines, $\phi$, can account for all of the observed dark matter by appropriately choosing a misalignment angle, $\Delta \theta \leq 1$. In the region below this band,  $\phi$ alone can constitute only a fraction of the dark matter even if we take the maximal value,  $\Delta \theta = 1$ (see dotted contours). The region above this band is excluded in our model because, $m_\phi f> v_\star^2$.  In blue we show the bounds from the MICROSOPE experiment on any apparent violation of the equivalence principle.}
		\label{fig:3}
	\end{figure} 

 The initial misalignment angle $\Delta \theta=\delta \phi_m/f$, can be varied by changing the inflationary Hubble scale (see \eq{misalign}). Given the requirement in \eq{req}, we find that the maximal value for the misalignment angle is $\Delta \theta \lesssim \epsilon$, where $\epsilon$ itself must satisfy $\epsilon\lesssim 1$. In Fig.~\ref{fig:3}  we show the `dark matter band', i.e. the region in the $m_\phi$-$f$ parameter space where $\phi$ can explain the dark matter density by adjusting $H_I$ to give an appropriate value of  $\Delta \theta$. This is the region between the two solid black lines in Fig.~\ref{fig:3}. We show the  value of $\Delta \theta$ required to obtain the observed dark matter density at the two edges of this band.    The region above the band is disallowed in our model because $m_\phi f= \epsilon \mu_\phi f$ cannot exceed $v^2$ by \eq{position}.  
 
 In the region below the dark matter band, $\phi$, can only account for a fraction of the observed dark matter even for a maximal, $\Delta \theta=1$. This is shown in Fig.~\ref{fig:3}  by contours of the maximal dark matter density, $\left[\rho_\phi\right]_{\rm max}$. In the presence of the couplings in \eq{lagphi}, the earth can source a spatial dependence of the amplitude of the  $\phi$-oscillations. This generates an additional force on the test bodies resulting in an apparent violation of the equivalence principle~\cite{Stadnik, gilad}. The strongest bounds on this effect were obtained by the MICROSCOPE experiment~\cite{microscope}; we show the corresponding constraints on our parameter space, assuming positive, ${\cal O}(1)$ values for the $c_i$ in \eq{lagphi} and  conservatively taking the misalignment angle  to be at its maximal value, $\Delta \theta=1$. The oscillations of $\phi$ would also lead to a time-variation of fundamental constants like the mass of electron and the fine-structure constants. We find, however, that present experiments (see for instance Ref.~\cite{poincare}) do not have the required sensitivity to probe our model mainly because the fraction of dark matter generated is not sufficient in the relevant parts of the parameter space.

\section{Conclusions}

\label{conclusions}

We have presented a solution to the hierarchy problem up to the scale, $\Lambda \sim \sqrt{v M_{pl}}$. Apart from the inflaton sector, we introduce a light PNGB scalar and an additional Higgs doublet. The main idea of our mechanism has been summarised in Fig.~\ref{fig:1} and~\ref{fig:2}. As shown in Fig.\ref{fig:1}, we consider a landscape of vacua across which the Higgs sector parameters, and thus the electroweak VEV, is allowed to vary. By construction, the vacuum energy of the Higgs sector peaks in vacua with an electroweak scale much smaller than the cutoff (see Fig.~\ref{fig:2}). During eternal inflation this results in patches with maximal vacuum energy---and a hierarchically small electroweak scale---expanding exponentially faster than other patches to eventually dominate the universe in volume. While this maximal energy state can still tunnel into lower vacua, such transitions are exponentially suppressed.

Our model builds on the earlier work of Ref.~\cite{Geller, Cheung, Giudicesol, Strumia}. As in these works, the key feature of our mechanism is the exponentially higher rate of inflationary expansion of patches where the electroweak VEV is small compared to the cutoff. There is, however, a crucial difference between our model compared to previous ones. In our model, the light field, $\phi$, does not scan the Higgs mass and plays a role only in the selection mechanism. We assume instead the presence of a landscape of vacua having different values for low energy parameters and thus different values of the electroweak VEV.  It is in this sense that our approach is minimal. While our model may be less ambitious in this regard, the decoupling of the scanning and selection mechanisms, in fact, gives our model a unique simplicity and several advantages.   The light field $\phi$,  in our model, has a completely generic PNGB-like potential with no additional small parameters. In particular, our model does not run into the issues discussed in Ref.~\cite{Gupta1} and thus does not require an elaborate clockwork sector. The field value of $\phi$ never exceeds its decay constant and is therefore never transplanckian. 

An appealing feature of this work is that, unlike the case of anthropic selection~\cite{Agrawal}, our selection mechanism is robust to the variation of almost all the parameters of our model (the only exception being the sign of the mass term in the $\phi$-potential). In particular, we freely vary the quartics of the 2HDM sector and the trigger term. We find that maximizing the vacuum energy with respect to the variation of these parameters automatically selects regions in the landscape with desirable properties---such as electromagnetism remaining unbroken after minimization of the 2HDM. 

The maximization of the vacuum energy with respect to the quartics also yields the smoking gun signature of our model, namely the prediction of $\tan \beta$ as a function of other measurable parameters, namely the heavier CP even Higgs mass, $m_H$ and the other mixing angle $\alpha$ (see \eq{prediction}). This is shown in Fig.~\ref{fig:4} where we show predictions for $\cos (\beta-\alpha)$---the parameter that controls the couplings of the SM-like Higgs boson---as a function of the mass of the heavier Higgs, $m_H$ and $\tan \beta$. This is a precise prediction that can be confirmed or falsified by careful measurement of the masses and couplings of the different Higgs bosons of the 2HDM sector.  As far as the pseudoscalar in our model is concerned, it can account for the observed dark matter density via the misalignment mechanism in certain regions of the parameter space (see Fig.~\ref{fig:3}). It can also be probed by experiments looking for a violation of the equivalence principle and future tests of the variation of fundamental constants.

To summarise we have proposed a simple, non-anthropic selection mechanism that explains the smallness of the weak scale and provides smoking gun signatures that can be probed in present and future experiments.

\paragraph{Acknowledgements} We are grateful to Sandip Trivedi for insightful discussions and comments that cleared up many of our questions about the landscape. We thank Abhishek Banerjee for patiently explaining to us the MICROSCOPE bounds and Siddhartha Karmakar for discussions during the initial stages of this work.  We also thank Ryuichiro Kitano and Avik Banerjee for their comments on this work. We acknowledge the support from the Department of Atomic Energy (DAE), Government of India, under Project Identification Number RTI 4002.

\appendix
\section{Quantum Fluctuations and the volume-weighted Fokker-Planck equation}
\label{FPV}
In this appendix, we discuss the role of quantum fluctuations of the scalar fields during inflation. We follow the approach and notation of Ref.~\cite{Graham, Cheung} closely.  We consider first the pseudoscalar field, $\phi$. During eternal inflation, $\phi$ undergoes quantum fluctuations which can be modeled as a random walk with a typical step-size of $\Delta \phi \sim H_I/2\pi$, superimposed on the classical slow-roll trajectory. This results in the evolution of the volume-weighted  distribution of $\phi$ across different Hubble patches, $P(\phi,t)$ according to the volume weighted Fokker-Planck (FPV) equation given by,
\begin{equation}\label{fpveq}
    \frac{\partial P}{\partial t} = \frac{\partial}{\partial \phi} \left[\frac{H_I^3(\phi)}{8 \pi^2}\frac{\partial P}{\partial \phi} + \frac{V'(\phi)}{3 H_I(\phi)}P \right] + 3H_I(\phi)P.
\end{equation}

Now using \eq{position} and \eq{hierarchy}, we see that the value of $V_\phi$ in \eq{vphi} (or indeed any of the terms of $V_H$ in \eq{vh}) is a small perturbation on the total vacuum energy during inflation. We thus substitute in \eq{fpveq}, $P(\phi,t) = \text{exp}(3H_{I_0}t)~\text{exp}(-\nu(\phi))~\psi(\phi,t)$, with $\nu(\phi) \equiv 4\pi^2 V(\phi)/3H_{I_0}^4$,  to obtain,
\begin{equation}
    -\frac{4\pi^2}{H_{I_0}^3} \frac{\partial \psi}{\partial t} = -\frac{1}{2} \frac{\partial^2 \psi}{\partial \phi^2} + \frac{1}{2}\left[-\nu''(\phi) + (\nu'(\phi))^2 - \frac{3}{M_{pl}^2} \nu(\phi)\right]\psi,
\end{equation}
which is the  Schr\"{o}dinger equation with $t \to i t$. Here, $H_{I_0}$ is the Hubble scale corresponding to the vacuum energy with  $\phi=0$. We can make an eigenvalue decomposition of the form $\psi(\phi,t) \equiv \sum_n c_n \psi_n(\phi) e^{-\Gamma_n t}$. The eigenstates  $\psi_n$ satisfy the time-independent Schr\"{o}dinger equation,
\begin{equation}\label{time-ind-schr-2}
    \frac{4\pi^2 f^2}{H_{I_0}^3} \Gamma_n \psi_n = -\frac{1}{2} \frac{\partial^2 \psi_n}{\partial x^2} + \frac{1}{2}\left[-\nu''(x) + (\nu'(x))^2 - \frac{3 f^2}{M_{pl}^2} \nu(x)\right]\psi_n,
\end{equation}
where  $x \equiv \phi/f$. After a sufficiently long time, it is expected that $\psi(\phi,t)$ will be dominated by the solution corresponding to the minimum eigenvalue which we call $\psi_0(x)$.

Taking $\nu(x) = \eta \sum_{n} \lambda_{n} x^{2 n}$, and $\eta \equiv 4\pi^2 \mu_\phi^2 f^2/3H_{I_0}^4$ we   estimate the different terms in   \eq{time-ind-schr-2} to be: $\nu'' \sim \eta$, $(\nu')^2 \sim \eta^2$ and $\frac{3 f^2}{M_{pl}^2} \nu(x) \sim \frac{f^2}{M_{pl}^2} \eta$. In the limit $\eta \gg 1$ (which gives the second equation in \eq{cbeatsq}),  the dominant contribution to the effective potential within the square brackets comes from the $(\nu')^2$ term which is minimized at the extrema of  $V_\phi$. Furthermore, assuming $f^2 \ll M^2_{pl}$ (which gives the third equation in \eq{cbeatsq}) we see that the $-\nu''$ term lifts the maxima of $V_\phi$ so that $\psi (x)$, and thus the volume weighted distribution,  peaks at the minima of $V_\phi$  as expected classically. 
 
 A similar analysis can be carried out for the Higgs fields. We can analogously define $\eta_H \equiv 4\pi^2 v^4/3H_{I_0}^4$ so that  $\eta_H \gg 1$ (which gives the first equation in \eq{cbeatsq})  becomes the limit in which the distribution function of the   Higgs fields are peaked around their classical minima.

We now estimate the spread of the distribution of the field, $\phi$. This determines the misalignment angle of $\phi$. If the dominant contribution to the potential comes from some power $p$ of $\phi$, i.e. $\nu(x) \sim \eta x^p$ and $\eta \gg 1$, then it is evident that if $\psi_{0,1}(x)$ is a solution of \eq{time-ind-schr-2} with $\eta = \eta_1$ and $\Gamma = \Gamma_{0,1}$, then $\psi_{0,2}(x) = \psi_{0,1}\left(\left(\frac{\eta_2}{\eta_1}\right)^{\frac{1}{p}}x\right)$ is the solution of \eq{time-ind-schr-2} with $\eta = \eta_2$ and $\Gamma_{0,2} = \Gamma_{0,1} \left(\frac{\eta_2}{\eta_1}\right)^{\frac{2}{p}}$. Thus, the width of $\psi_0$ must scale as  $\eta^{-\frac{1}{p}}$. The width of $P(\phi,t)$ also goes as  $\eta^{-\frac{1}{p}}$ because the argument of the exponential $\text{exp}(-\nu(\phi))$ is invariant under such a rescaling. This gives us the misalignment scale given by \eq{misalign}. Note that for  the quadratic regime discussed in Sec.~\ref{phipheno}, we must take $\eta_{q} \equiv 4\pi^2 m_\phi^2 (\delta \phi_q)^2/3H_{I_0}^4$ and $\delta \phi_m\sim \delta \phi_q/\eta_q$.

 \section{Minimization of the 2HDM potential}
 \label{2hdm}
 In this appendix, we provide details about the minimization of the 2HDM potential in \eq{2hdmpot} using some of the results of Ref.~\cite{diazcruz}. The  8 minimization conditions corresponding to the 8 components of the two doublets, 
 \begin{equation}\label{components}
 H_1 = \frac{1}{\sqrt{2}}\begin{pmatrix}
     \phi_1 + i \phi_2\\
     \phi_3 + i \phi_4
 \end{pmatrix} ~~~~~~H_2 = \frac{1}{\sqrt{2}} \begin{pmatrix}
     \phi_5 + i \phi_6\\
     \phi_7 + i \phi_8
 \end{pmatrix},
 \end{equation}
 are given by,
 \begingroup
 \allowdisplaybreaks
 \begin{align}
     \phi_1:&~~ \frac{u}{\sqrt{2}} v_1 v_2 \left(\lambda_4~\text{cos} \xi + \hat{\lambda}_5~ \text{cos}(\alpha + \xi)\right) = 0\\
     \phi_2:&~~ \frac{u}{\sqrt{2}} v_1 v_2 \left(-\lambda_4~\text{sin} \xi + \hat{\lambda}_5~\text{sin}(\alpha + \xi)\right) = 0\\
     \phi_3:&~~ \frac{v_1}{\sqrt{2}}\left(2 \lambda_1 v_1^2 + u^2 \lambda_3 + v_2^2 (\lambda_3 + \lambda_4) + 2 \mu_1^2 + \hat{\lambda}_5~\text{cos}(\alpha + 2 \xi)~v_2^2\right) = 0\\
     \phi_4:&~~ \frac{v_2^2}{\sqrt{2}} v_1 \hat{\lambda}_5~ \text{sin}(\alpha+ 2 \xi) = 0\\
     \phi_5:&~~ \frac{u}{\sqrt{2}} \left(2(u^2+v_2^2) \lambda_2 + v_1^2 \lambda_3 + 2 \mu_2^2\right) = 0\\
     \phi_6:&~~ 0\\
     \phi_7:&~~ \frac{v_2}{\sqrt{2}} \left(2(u^2+v_2^2) \lambda_2 + v_1^2 (\lambda_3+\lambda_4) + \mu_2^2 \right)\text{cos}\xi + \frac{v_1^2}{\sqrt{2}} v_2 \hat{\lambda}_5~ \text{cos} (\alpha+\xi) = 0\\
     \phi_8:&~~ \frac{v_2}{\sqrt{2}} \left(2(u^2+v_2^2)\lambda_2 + v_1^2 (\lambda_3 + \lambda_4) + 2 \mu_2^2 \right)\text{sin} \xi - \frac{v_1^2}{\sqrt{2}} v_2 \hat{\lambda}_5~ \text{sin}(\alpha + \xi) = 0.
 \end{align}
 \endgroup
where we have used $SU(2)_L \times U(1)_Y$ invariance to write, 
\begin{equation}\label{genvev}
  \langle H_1 \rangle = \frac{1}{\sqrt{2}}\begin{pmatrix}
    0\\
    v_1
\end{pmatrix}~~~~~~~~~~
\langle H_2 \rangle = \frac{1}{\sqrt{2}}\begin{pmatrix}
    u\\
    v_2 e^{i \xi}
\end{pmatrix}.
\end{equation}

 From the above equations, it is evident that the VEVs of the two doublets must take the form in \eq{vev1} and \eq{vev2}. When $u = 0$ the doublet VEVs are given by  \eq{vev1} and minimization with respect to $\xi$ yields two solutions, $\xi=-\alpha/2, -\alpha/2 +\pi$ but only the first one is relevant to us as the second solution does not satisfy the condition in \eq{stability} as we have chosen $\kappa>0$. Thus, the potential for this case after substituting \eq{vev1} becomes,
\begin{equation}
    V_{\text{2HDM}} = \frac{\mu_1^2 v_1^2}{2} + \frac{\mu_2^2 v_2^2}{2} + \frac{\lambda_1 v_1^4}{4} + \frac{\lambda_2 v_2^4}{4} + \frac{\lambda_{345}~v_1^2 v_2^2}{4} \label{2hdm_vev1},
\end{equation}
 where we recall that $\lambda_{345} = \lambda_3 + \lambda_4 + \hat{\lambda}_5$.
Minimizing \eq{2hdm_vev1} with respect to $v_1$ and $v_2$, we obtain
\begin{equation}\label{muvev}
    \mu_{1,2}^2 = - \lambda_{1,2} v_{1,2}^2 - \frac{\lambda_{345}}{2} v_{2,1}^2 \; ,
\end{equation}
which can be used to obtain \eq{vac2}. We can also use \eq{muvev} to write  $v$ and $\text{tan}\beta$ in terms of the underlying parameters of the potential,
\begin{align}\label{muvev2}
    v^2 &= \frac{4 \lambda_2 \mu_1^2 + 4 \lambda_1 \mu_2^2 -2 \lambda_{345} (\mu_1^2 + \mu_2^2)}{\lambda_{345}^2 - 4\lambda_1\lambda_2}\nonumber\\
    \text{tan}^2\beta &= \frac{2 \lambda_1 \mu_2^2 - \lambda_{345} \mu_1^2}{2 \lambda_2 \mu_1^2 - \lambda_{345} \mu_2^2}.
\end{align}

We now show that class II minima always have smaller vacuum energy than class I minima. For this we consider the potential $\hat{V}_{2HDM}$ and study its minimisation   in the radial direction, $\rho=\sqrt{\sum_{i=1}^8 \phi_i^2}$. First, we use  \eq{genvev} to express the two doublets as,
\begin{equation}
  \langle H_1 \rangle = \frac{1}{\sqrt{2}}\begin{pmatrix}
    0\\
     \rho~ \text{cos} \beta
\end{pmatrix}~~~~~~~~~~
\langle H_2 \rangle = \frac{1}{\sqrt{2}}\begin{pmatrix}
    \rho~ \text{sin} \beta\\
   \rho~ \text{sin} \beta~ \text{cos} \gamma~e^{i \xi}
\end{pmatrix}.
\end{equation}
which can be used to write the potential as follows,
\begin{equation}\label{radialform}
    \hat{V}_{2HDM} =\frac{\hat{\mu}_\rho^2 \rho^2 }{2} +\frac{\hat{\lambda}_\rho \rho^4}{4}.
\end{equation}
where,
\begin{align}
   \hat{\lambda}_\rho \equiv& \biggl(\lambda_1~\text{cos}^4 \beta + \lambda_2~\text{sin}^4\beta + \frac{\lambda_3}{4}~\text{sin}^2 2\beta+ \frac{\lambda_4}{4}~\text{sin}^2 2\beta~\text{cos}^2\gamma + \frac{\hat{\lambda}_5}{4}~\text{sin}^2 2\beta~\text{cos}^2\gamma~ \text{cos}(\alpha + 2\xi)\nonumber\\ 
   & \;\; - \frac{\kappa^2}{16\lambda_\phi}~\text{sin}^2 2\beta~ \text{cos}^2\gamma ~\text{cos}^2\xi \biggr),\\
    \hat{\mu}_\rho^2 \equiv& \left(\mu_1^2~\text{cos}^2\beta + \mu_2^2~\text{sin}^2\beta + \frac{\kappa \mu_\phi f}{8\lambda_\phi}~\text{sin}2\beta~\text{cos}\gamma~\text{cos}\xi\right).
\end{align}
Using the minimization conditions we get, 
\begin{equation}
v^2 =\frac{-\hat{\mu}_\rho^2}{\hat{\lambda}_\rho}
\end{equation}
where the vacuum energy  contribution from the 2HDM sector, 
\begin{equation}
{\cal VE}_{2HDM}=\frac{-\hat{\mu}_\rho^4}{4 \hat{\lambda}_\rho}
\end{equation}
is clearly negative.

Finally, let us show that class I and III minima cannot coexist. We can again recast the $\phi$-independent part of the potential in \eq{vh},  $V_{2HDM}$, to the form given in  \eq{radialform}, i.e., 
 \begin{equation}\label{radialform2}
    V_{2HDM} =\frac{\mu_\rho^2 \rho^2}{2}  +\frac{\lambda_\rho  \rho^4}{4}.
\end{equation}
where $\left\{\mu_\rho, \lambda_\rho\right\}=\left\{\mu_\rho, \lambda_\rho\right\}\vert_{\kappa \to 0} $. This potential will have an electroweak symmetry breaking minima only if $\mu_\rho^2<0$ and $\lambda_\rho^2>0$. The first of these conditions can be written as
\begin{equation}
    \mu_1^2 c_\beta^2+\mu_2^2 s_\beta^2 <0
\end{equation}
which requires either $\mu_1^2$ or $\mu_2^2$  to be negative. Thus class III minima cannot coexist with class I minima where both $\mu_1^2$ and $\mu_2^2$ must be positive. We can also use \eq{radialform2} to write the vacuum energy at the minimum to be $-\lambda_\rho v^4/4$ which again gives \eq{vac2}.

\providecommand{\href}[2]{#2}\begingroup\raggedright\endgroup

\end{document}